# Synthesis and Electrochemical Study of Multi-Phase, Multi-Species Ion Conductor Sodium $\beta''$-Alumina (BASE) + 20SDC Using a Vapor-Phase Process

Pooya Elahi, Jude A. Horsley, Taylor D. Sparks

Materials Science and Engineering, University of Utah, UT 84112, USA

**Abstract**

The recent emergence of multi-species multi-phase materials provides intriguing opportunities to maximize electrochemical performance in various electrochemical devices. This work summarizes the current understanding of the coupled transport reactions in multi-phase multi-species ionic conductors. We also provide experimental results of fabrication of multi-phases $Na - \beta'' - alumina + 20\ mol\%\ Scandia\ Doped\ Ceria\ (20SDC)$ as a simultaneous $Na^+$ and $O^{2-}$ ion conductor by a cost-effective vapor phase process demonstrating higher conductivity achieve in a much shorter time, as compared to other published results. In this study, two-phase contiguous composites of $\alpha - Al_2O_3 + 20SDC$ are fabricated by conventional ceramic processing and sintering in air at $1400\ °C$, $1500\ °C$, and $1600\ °C$ for 3 hours. The samples are heat-treated exposed to a $Na_2O$ vapor source at different time lengths. The conversion mechanism involves coupled transport of sodium ions through newly formed $Na - \beta'' - alumina$ and oxygen ions through $SDC$. The experimental data are analyzed using diffraction and spectroscopy methods. The samples with finer grains show faster kinetics compared to coarse microstructures due to the presence of more extended triple-phase boundaries (TPB). As a result, the total conductivity of the multi-phase sample compared to that of pure 20SDC is improved by three times, while fabrication time is decreased by 60% comparing to $Na - \beta'' - alumina + YSZ$.

**Keywords**: Solid-state electrolytes, $Na - \beta'' - alumina$, $20SDC$, multi-phase multi-species ionic conductors, sodium-ion batteries.

**Introduction**

The vast majority of solid-state ionic conductors studied in the literature are typically capable of single species transport only. For example, yttria-stabilized zirconia (YSZ) [1–3]; rare-earth oxide-doped ceria [4–6] such as samaria-doped ceria (SDC), scandia-doped ceria (ScDC), or gadolinia-doped ceria (GDC); yttria-doped bismuth oxide (YDB) [7], and magnesium-doped lanthanum gallate (LSGM) [8,9] are all excellent oxygen ion conductors. Lithium lanthanum zirconium oxide (LLZO) and LiSICON are lithium superionic conductors [10]. Sodium beta and beta double prime alumina ($Na - \beta/\beta'' - alumina$, aka BASE or SBA) [11–14], Nafion [15,16], and NASICON [17,18] are excellent sodium ion conductors.

Another class of ionic conductors is single-phase materials that are nonetheless capable of multi-species ionic conduction. For instance, it has been shown that the addition of rare earth oxides as a dopant to alkaline earth zirconates, as well as cerates, will result in protonic $[H^+]$, hole $[h^\bullet]$, and oxygen $[O^{2-}]$ ionic conductivity in a single-phase material. Mixed Ionic and Electronic Conductors (MIECs) are other examples of a single-phase but multi-species conductor. MIECs simultaneously accommodate both ionic conductivity alongside electronic transport, making them a good candidate for membrane reactors in active electrochemical devices such as solid-state batteries, fuel cells, and electrochemical sensors [19–22]. Enhancement in electronic conductivity is common for these materials. Theoretically and experimentally, the local equilibrium thermodynamics implies that higher electronic concentration ($[e']$ and $[h^\bullet]$) decreases the permeate species' local chemical potential. The small but immensely effective higher electronic conductivity will be favorable for preventing electrolyte degradation, such as delamination of the electrolyte adjacent to the oxygen electrode in solid oxide electrolyzer cells (SOECs) [23–28]. Yttrium-doped barium zirconate ($Y - BaZrO_3$) and gadolinium-doped barium cerate ($Gd - $

$BaCeO_3$) are examples of single-phase but multi-species ionic conductor materials where proton $[H^+]$, oxygen $[O^{2-}]$, and electronic defects $[h^\bullet]$ are the mobile species [29,30].

Although the multi-species transport of charged species is provided, it is not thermodynamically possible to tune the total conductivity for all species simultaneously. For instance, enhancing protonic conductivity through addition of $H_2O$ content in the atmosphere simultaneously decreases the oxygen ion conductivity and electronic defects transport [31]. Such perovskites of the general type $Z$-doped $ABO_3$, when $Z$ is a lower valence ion substituting $B^{3+,4+}$ sites, can exhibit a dual electrochemical performance affected by operating conditions. In the dry air (absence of water-vapor), the facile rotation of corner-sharing oxygen octahedra exhibited by perovskite structures enables superior oxygen and electronic transport [19,20]. However, if heated in air in the presence of $H_2O$, the structure absorbs $H_2O$ molecules into the lattice, and it becomes a proton conductor [32,33].

An emerging third class of ionic conductors has been recently established where a composite consisting of multiple phases is able to accomplish uncoupled simultaneous multi-species ionic conduction. A notable example of such composite is the mixture of $Na - \beta'' - alumina + YSZ$ [34], where oxygen ion conduction transpires in the $YSZ$ phase while sodium ion conduction occurs in the BASE phase. The conductivity, alongside other materials properties of the composite, can be tuned by altering the ratio and distribution of the different phases [35]. While it is expected that the conductivity of a composite would underperform that of a single-phase material, this is not necessarily true as it depends on the sign of the Onsager transport coefficients in non-equilibrium linear thermodynamics [36–38]. Moreover, there could be other benefits such as an increased number of mobile species for enhanced electrochemical performance and strengthening, which was observed for other systems such as $Na - \beta'' - alumina + YSZ$ composites [39–43].

While there has been growing interest in development of new multi-species multi-phase ionic conductors, there are significant unanswered problems with these materials. For example, in the vapor phase conversion of $Na - \beta'' - alumina + 3$ or $8YSZ$, $3YSZ$ resulted in better mechanical properties compared to $8YSZ$. However, $3YSZ$ shows significantly lower ionic conductivity compared to $8YSZ$, thus longer conversion times. On the other hand, $8YSZ$ results in larger grain size, which decreases the TPBs, therefore, increasing the conversion time, and limiting the homogeneity and contiguity of the composite resulting in $\alpha - Al_2O_3$ remnant [44].

In this work we attempted to bypass currents issues with the vapor phase conversion of $BASE + YSZ$ phases such as long conversion time in an energy-intensive process, destabilization of the YSZ due to the unwanted transformation of fluorite to tetragonal and monoclinic phases during conversion, and poor mechanical properties. We show the possibilities of fabrication of a dual-phase, dual-species intermediate-temperature ionic conductor consisting of the two contiguous mixtures of $Na - \beta'' - alumina$ and $20\ mol\%$ samaria-doped ceria ($20SDC$) as sodium and oxygen ion conductors, respectively. In our study, improved mechanical properties for the converted sample are achieved, while the conversion time is decreased by almost 60%, and total conductivity of the composite increased comparing to other published data. Hence, two goals of this work achieved: 1) the fast fabrication of BASE and 2) fabrication of a sodium and oxygen ionic conductor using a composite ionic conductor to create the potential for designing unprecedented electrochemical cells with self-dependent multi-species transport through the electrolyte. Fabricated electrolyte was characterized physically and electrochemically using diffraction, electron microscopy, and spectroscopy techniques. This process showed the faster conversion time while the total conductivity of the sample was enhanced compared to that of pure $20SDC$ and previous published data on BASE+3YSZ.

**Experimental**

A contiguous mixture of $\alpha - Al_2O_3$ (Baikowki) and $20\ mol\%$ samaria-doped ceria ($20SDC$) (Fuelcellmaterials) powder was wet-milled in a weight ratio of $7:3$ in ethanol (200 proof) at 300 rpm for 10 hours using a planetary milling machine (Fritsch Pulverisette). After drying the slurry overnight at room temperature, powders were collected and sieved #70 (ASTM E11). The mixed powder was pelletized using a uniaxial press (Carver) at $110\ MPa$ followed by cold isostatic pressure at $220\ MPa$. The pellets were sintered at $1400\ °C$, $1500\ °C$, $1600\ °C$, for 3 hours in air using a chamber furnace (Carbolite Gero). The density measurement was carried out using the gas pycnometry method (Ultrapyc Anton Paar) using helium as the gas medium. The samples with relative density of $D_{rel} > 95\%$ were collected to continue the conversion process

The sintered pellets were placed in an alumina crucible loosely packed with $Na - \beta'' - alumina$ powder covered with the lid. The $Na - \beta'' - alumina$ powder was prepared using the zeta method can be found elsewhere [45], containing $\sim 8.85\ wt\%\ Na_2O$, $90.4\ wt\%\ Al_2O_3$, and $0.75\ wt\%\ Li_2O$ [46]. The covered crucible was placed in a furnace for heat treatment at $1250\ °C$ for different durations with intervals of 5 hours.

After the conversion, samples were cut using a high precision diamond saw (Allied High Tech) fine polished to 200 nm (Buehler) and thermally etched in the air at $1100\ °C$ for 1 hour. The microstructure of samples was investigated using cross-sectional SEM techniques and elemental mapping of the samples using EDS (FEI Quanta 600 FEG). The grain size was measured using the mean linear intercept method using ImageJ [47]. X-ray diffraction (Bruker D2 Phaser) with $Cu, K\alpha = 1.5406\ Å$, $2\theta = 10\ to\ 80\ degree$ was carried out before and after the conversion to ensure the successful conversion throughout the bulk sample. The diffraction pattern peaks were obtained from 10 to 80 $2\theta$ degree, and indicating the presence of $\alpha - Al_2O_3$ and $20SDC$ without the presence of any secondary phase or contaminations after sintering the pellets. Rietveld refinement was conducted using full pattern Rietveld method based on least-squares refinement using GSAS-II software [48]. Porous Platinum electrodes (Heraeus) were screen printed symmetrically on both sides of the samples ($\sim 10\ \mu m$) and fired at $950\ °C$ for $1h$. The Electrochemical Impedance Spectroscopy (EIS) (Solartron - Impedance/Gain-Phaser Analyzer SI 1260 & Electrochemical interface SI 1287) was performed on converted samples and pure $20SDC$ with a frequency range of $1\ MHz$ to $1\ Hz$, with $10\ mV$ amplitude from $200\ °C$ to $800\ °C$. The exact temperature of the sample was monitored using a $K$-type thermocouple (Omega) and a Keithley 2000 meter.

**Results and Discussion**

Figure 1 shows the conversion thickness of the samples versus exposure time to $Na_2O$ source. The samples with finer microstructure exhibit higher conversion thickness at the same time as that of coarse-grained structures, which is consistent with the theoretical model presented by Parthasarathy [49]. Thus, finer grains facilitated the conversion process. The conversion time is considered as the dwell time once the conversion temperature is achieved inside the furnace ($1250\ °C$). Therefore, this should be why extrapolation of the polynomial fit to the data does not occur at $t = 0\ h$. Due to the high partial vapor pressure of the $Na_2O$, before reaching $1250\ °C$, some sodium oxide vapor was provided inside the covered crucible, which made the reaction propagate forward. The conversion of the samples can be explained in two steps. In step one, a thin layer of $Na - \beta'' - alumina$ formed on the pellet's surface (initiated below $1250\ °C$), and later at sufficiently high temperatures, $O^{2-}$ conduction occurs through $20SDC$ at the three-phase boundaries (TPB). Since TPB's length is inversely proportional to the grain size, a decrease in grain size (d) results in longer TPB, shorter diffusion distance, and, subsequently, higher conversion thickness.

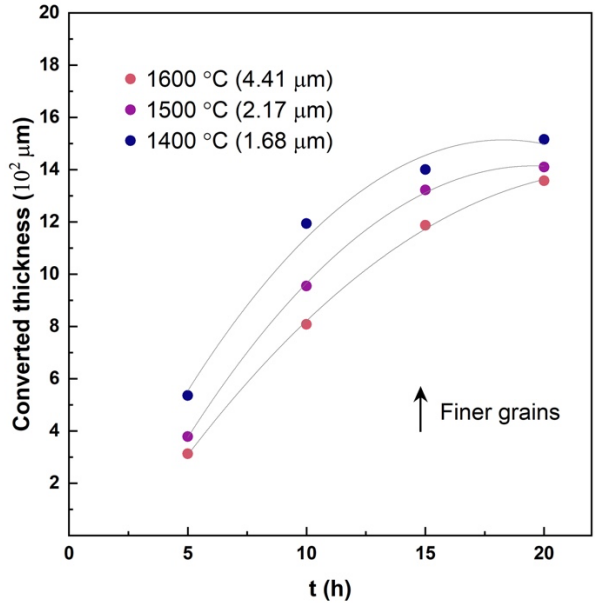

*Figure 1. Converted thickness as a function of the dwell time with the polynomial fit at* $1250\ °C$, *for samples sintered at* $1400\ °C$, $1500\ °C$, *and* $1600\ °C$, *with the average grain size of* $1.68$, $2.17$, *and* $4.41$, *respectively.*

Figures 2 (a) and 2 (b) show the XRD patterns of the as-sintered sample before conversion ($Al_2O_3 + 20SDC$) and fully converted sample ($Na - \beta'' - alumina + 20SDC$). Figure 2 (a) shows the XRD pattern of the as-sintered sample at $1600\ °C$ for 3 hours. The $20SDC$ crystal structure has a cubic crystal structure of $CeO_2$ and Miller indices of the first six peaks are (111), (200), (220), (311), (222), (400), and (331). However, $CeO_2$ peaks are slightly shifted to the left due to the presence of $20\ mol\%$ of $Sm_2O_3$ akin to $Sm_{0.2}Ce_{0.8}O_{1.9}$ in the crystallography open database (COD 4343142) [50]. The lattice parameter of the $20SDC$ was calculated to be $5.4745$ Å which is a good agreement with previous studies [51]. Crystallite size was reported to be $13.82\ nm$, using Scherrer's equation. The $\alpha - Al_2O_3$ phase is characterized as a hexagonal crystal structure (COD 1010914). The first six peaks are labeled as (012), (104), (110), (113), (024), and (116). The lattice parameter is $a = b = 4.7514$ Å and $C = 12.9621$ Å. Figure 2 (b) shows the presence of the $Na - \beta'' - alumina$ (COD 1529826) and $20SDC$ as labeled. $20SDC$ doesn't show any significant change in lattice parameters and peaks positions. The lattice parameter of $Na - \beta'' - alumina$ is $a = b = 5.6122$ Å and $c = 33.7518$ Å. Comparing the before and after conversion XRD pattern, the prominent corundum peaks have been eliminated and replaced by $Na - \beta'' - alumina$ peaks, which indicates the complete conversion of the $Al_2O_3$ to $\beta''$-alumina and formation of a contiguous dual-phase.

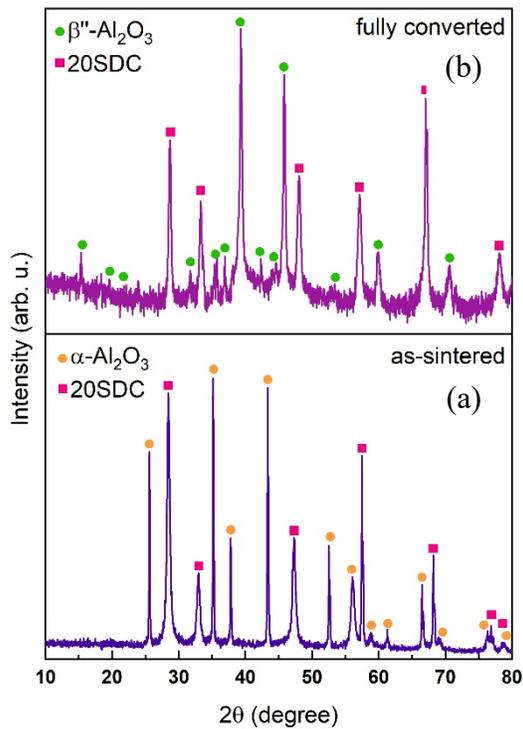

*Figure 2. XRD pattern of the samples, (a) before and (b) after full conversion.*

Figure 3 represents the SEM micrographs of the $Al_2O_3 + 20SDC$ partially converted samples to $Na - \beta'' - alumina + 20SDC$ after 5 hours of heat treatment at $1250\ °C$. Figure 3 (a) shows the cross-sectional SEM of partially converted samples after 5 hours at $1250\ °C$. The converted thickness measured ~350 μm. Region A shows the converted $Al_2O_3 + 20SDC$ to $Na-\beta''-Al_2O_3$, while region B is the pristine, dense unconverted $Al_2O_3 + 20SDC$ region. From the SEM investigated cross-section, the conversion took place from the surface through the sample bulk by forming $Na-\beta''-alumina + 20SDC$ toward the interface band.

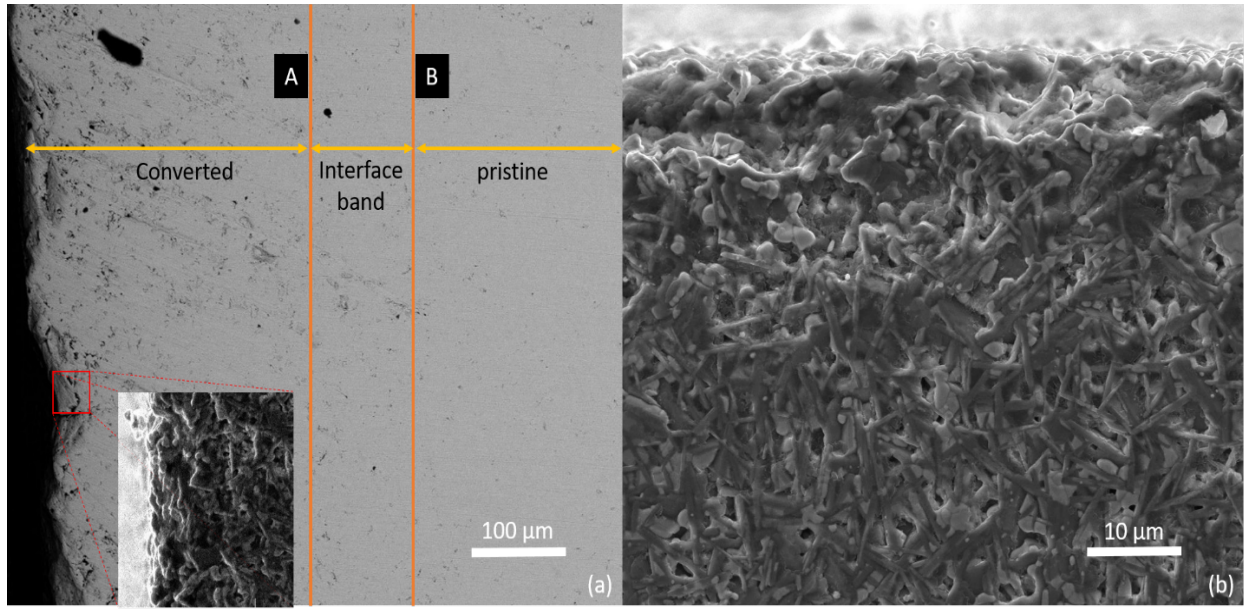

Figure 3. Cross-sectional SEM micrograph of the partially converted sample at 1250 °C for 5 hours, including (a) the converted region, interface band, and pristine sample bulk and (b) the converted edge at the high magnification. The conversion thickness is ~350 μm.

Figure 3 (b) exhibits the microstructure of the converted region of the sample. The change in the shape of the grains indicates the formation of the new phase at the surface compared to the bulk of the sample at the central region. According to the phase diagram of $Na_2O - Al_2O_3$, the conversion temperature could be chosen in a reasonably broad range from ~1200 °C to 1500 °C.

In the current work, $T = 1250\ °C$ was chosen as the heat treatment temperature to avoid any liquid phase formation possibly close to the eutectic reaction occurring around ~1550 °C, to optimize the kinetics and thermodynamics of the reactions to minimum loss of $Na_2O$ from the packing powder during the vapor phase conversion while conversion reaction propagates. The interface propagation happens in a fairly heterogenous manner. The conversion interface exhibits some microstructural defects while the reaction between $Al_2O_3$ and $Na_2O$ gas takes place, which results in a significant volume change.

The sample sintered at $1600\ °C$ shows a high relative density ($> 95\%$) with few pores confined at the $3-$ and $4-$grain boundaries as well as pores trapped in the grains. The presence of trapped pores probably happened by grain boundaries that swept past pores during the sintering and microstructure evolution. However, the converted region "A" has some porosity. The origin of the newly formed pores in the converted region is attributed to significant volume change while $Al_2O_3$ is exposed to $Na_2O$ to form $Na-\beta''-alumina$. Figure 3 (b) shows a higher magnification of the converted region.

Figure 4 represents the SEM micrograph of fully converted samples. The darker grains are $Na-\beta''-alumina$ and the lighter grains are $20SDC$ grains. The micrograph of the analyzed region shows the formation of the typical prismatic $Na-\beta''-alumina$ grains at the expense of the $Al_2O_3$ particles and the formation of some voids as a common occurrence.

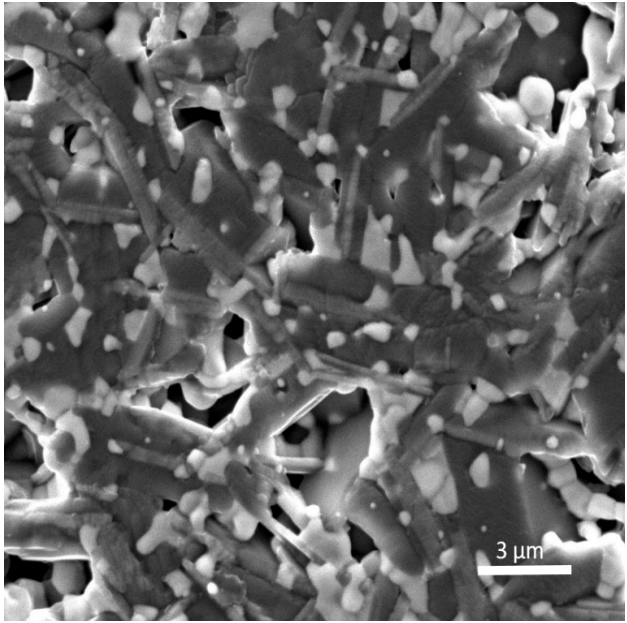

Figure 4. SEM micrograph of the converted structure after thermal etching at $100\ °C$ below sintering temperature, for 1 hour. The prismatic shape of $Na-\beta''-alumina$ with the presence of dispersed $20SDC$ can be observed. Due to the volume change after the conversion of $Al_2O_3$ to $Na-\beta''-alumina$, a number of pores are present in the micrograph.

Figure 5 (a-d) demonstrates the 2D elemental mapping of $5x5\ \mu m$ converted area, which endorses the formation of the $Na - \beta'' - alumina$ as a complementary technique to characterize the presence of sodium in the structure.

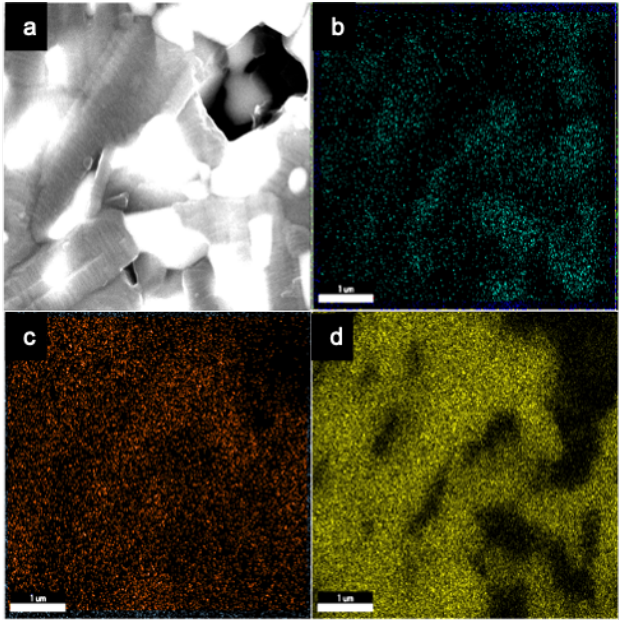

*Figure 5. (a) SEM micrograph of the investigated area, and 2D elemental mapping of the converted cross-section for (b) cerium, (c) sodium, (d) aluminum. The homogenous sodium distribution ensured the samples' full conversion without residual $Al_2O_3$.*

Figure 6 shows Electrochemical Impedance Spectroscopy (EIS) spectra of the converted samples after full conversion. The data was obtained from $1MHz$ to $1Hz$. Figure 6 (a) shows the full EIS spectra in the low-temperature regime ($200\ °C$ to $300\ °C$), and figure 6 (b) corresponds to the high-frequency regime at the same temperature range. Figure 6 (c) shows the full EIS spectra in the high-temperature regime ($540\ °C$ to $570\ °C$), and figure 6 (d) corresponds to the high-frequency regime at the same temperatures. The EIS conditions such as electrode material, sintering temperature, sample thickness, and testing environment were constant for all samples. It has been proven that a significant effect resulting from inductive load from the leads and instrument contributes to the EIS data, especially at a high-frequency regime. Therefore, it is crucial to subtract the inductive load effect from the obtained data to eliminate the effect of instrumentation and leads [52]. This inductance appears as an imaginary part of the impedance data in the positive direction of the imaginary part of the impedance axis. Notably, there is a non-considerable dependence of inductance on temperature, unlike the sample's behavior.

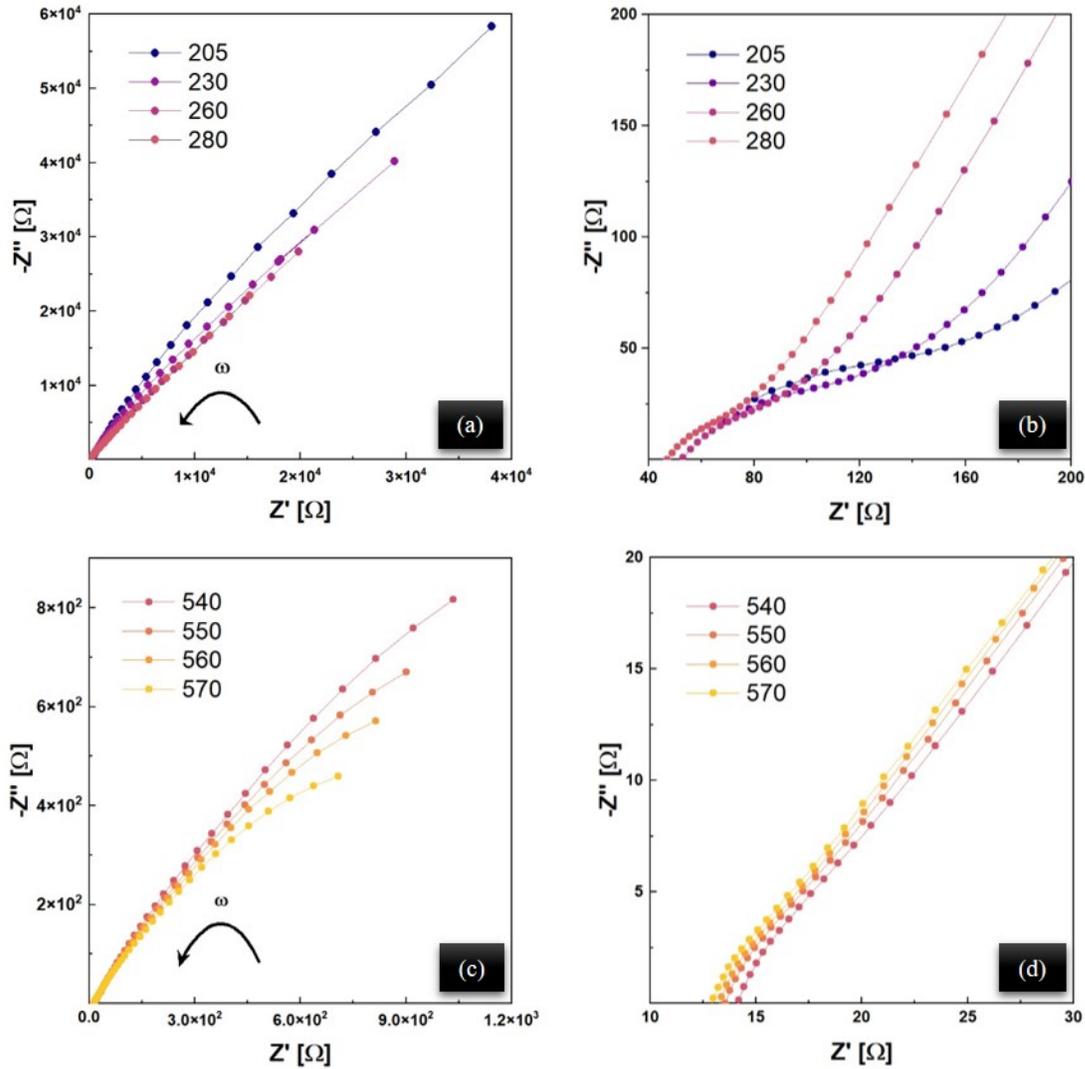

*Figure 6. EIS spectra of the low and high-temperature regime of converted samples in (a & c) frequency range from 1MHz to 1Hz, and (b & d) enlarged high-frequency regime. Arrow indicates the direction of increase in angular frequency.*

The extrapolation of the data to the x-axis ($Z'$ real part of complex impedance) was considered as measured. Figure 7 shows the Arrhenius plot of the total conductivity. The data points were fit to the conductivity equation

$$\sigma = \left(\frac{\sigma_0}{T}\right)\exp\left(-\frac{E_a}{k_B T}\right)$$

where $\sigma_0$ is the pre-exponential factor in $S.cm^{-1}.K$, T is the temperature in $K$, $E_a$ is the activation energy, and $\kappa_B$ is the Boltzmann constant. The values of activation energy and pre-exponential factor have been obtained from the linear fits to the data presented in Figure 7 and calculated to be $E_a = 0.23\ eV$, and $\sigma_0 = 221.4\ S.cm^{-1}.K$ for composite $Na - \beta" - alumina + 20SDC$ and $E_a = 0.68\ eV$, and $\sigma_0 = 80{,}017.45\ S.cm^{-1}.K$ for pure $20SDC$. The measured and calculated activation energy for pure $20SDC$ as the reference is in good agreement with Mansilla et al. [53], where they reported $E_a = 0.6\ eV$.

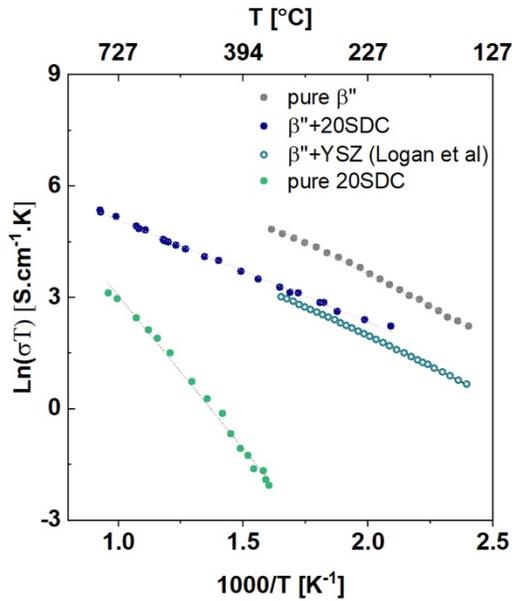

Figure 7. Arrhenius plot of the measured conductivity on fully converted and pure 20SDC samples.

Microstructure plays a crucial role in conductivity measurements. Fine-grained structures result in lower conductivity owing to resistance associated with lower transport and higher scattering of charged species at the grain boundaries. Coarse-grained structures, however, typically result in higher conductivity. In other words, grains are less resistive in comparison with grain boundaries.

The equivalent circuit for a two-phase, two species ionic conductor was suggested based on the resistance and capacitance of each phase and electrode effect. The electrode resistance and capacitance contribute to low-frequency data. At intermediate frequencies, the electrode effect fades away. At sufficiently high frequency, the capacitance of the phases SDC and $\beta''$ is shorted as typical behavior of ionic conductors. Hence, as presented in figure 8 (a) the equivalent circuit will be simplified as illustrated in figure 8 (b).

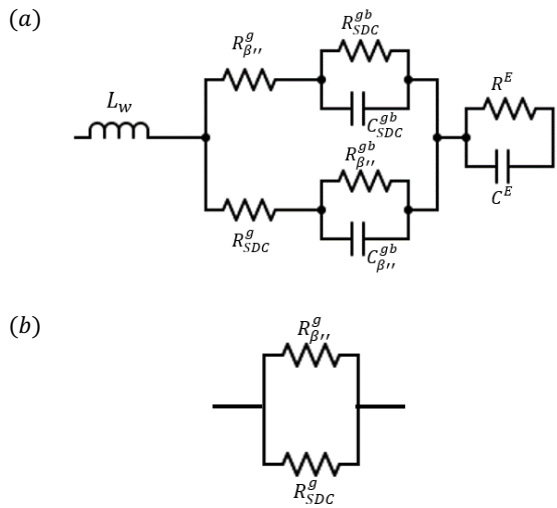

Figure 8. *Suggested equivalent circuit describing transport through* $Na-\beta''-alumina+20SDC$ *composite. At high frequencies after subtraction of induction load, the equivalent circuit will behave as (b), where* $L_w$ *is the lead induction and* $R_i^g, R_i^{gb}, C_i^g, C_i^{gb}$ *are the grain resistance, the grain boundary resistance, grain boundary capacitance in phase* i, *respectively.* $R^E$ *and* $C^E$ *represent the electrode resistance and electrode capacitance.*

At high frequency regime, grain and grain boundary capacitances are shorted. Also, the electrode parameters $(R^E, C^E)$ are not reflected in high frequency regime. Therefore, after subtraction of induction load from the data, at high frequency intercepts, measured resistance can be given by

$$R_{high\ freq.} = \frac{R_{SDC}^g\ R_{\beta''}^g}{R_{SDC}^g + R_{\beta''}^g}$$

$Na-\beta''-alumina$ and $20SDC$ both are well-known intermediate-temperature ionic conductors, however at lower temperatures $R_{\beta''}^g < R_{SDC}^g$ and at higher temperatures $R_{\beta''}^g > R_{SDC}^g$.

Therefore, the higher conductivity of the $Na - \beta'' - alumina$ especially at lower temperatures contributed to the increase in total conductivity of the dual-phase sample and caused an increase in total conductivity compared to the pure $20SDC$ sample, as data presented in figure 7.

In the current study, the conductivity behavior of the sample was found to be linear in the Arrhenius plot, which is attributed to the convolution of both phases' conductivity at lower temperatures. In future work, authors will directly add $Li_2O,$ and $MgO$ to basic starting mixture to increase the total conductivity of converted samples to introduce more $Na^+$ ions into the conduction planes, using higher $Na_2O$ content and larger grain size, at the expense of the higher conversion time to remove the trace of alumina and make pure $Na - \beta'' - alumina$ .

**Summary**

In the present work, dual-phase, dual-species (sodium and oxygen ions) ionic conductors were fabricated by vapor phase conversion. Two-phase composites consist of $Al_2O_3$ and $20SDC$ were mixed as a contiguous mixture and were exposed to $Na_2O$ source. Samples with the thickness of $1.2\ mm$ after about 10 hours could be converted to $Na - \beta'' - alumina + 20SDC$. There was no sign of reaction between $20SDC$ and $Na_2O$ and $20SDC$ remained the same throughout the conversion process as investigated by X-ray diffraction. The microstructure characterization showed that although samples sintered at lower temperatures, had finer grains, and consequently faster conversion times, lowering the sintering temperature resulted in lower conductivity for the converted sample due to the higher ion scattering at the extended TPBs and grain boundaries. The conductivity measurement carried out using EIS indicated the increase in total conductivity compared to pure $20SDC$ and previous works using $YSZ$. Provided that one makes two electrodes with different chemical potential of sodium and oxygen ions, the novel multi-phase, multi-species ionic conductor $Na - \beta'' - alumina + 20SDC$ could be used as a novel solid-state electrolyte for active electrochemical devices such as fuel cells and pumps and sensors which suffice simultaneous transport of multi-species, thus high and tunable properties such as ionic and electronic conductivity while enhancing mechanical, thermal and performances. The use of $20SDC$ decreased the conversion time by almost 60% while providing higher oxygen ion conductivity and better mechanical properties.